\documentclass[%
 reprint,
superscriptaddress,
 amsmath,amssymb,
 aps,
]{revtex4-1}

\usepackage{graphicx}
\usepackage{dcolumn}
\usepackage{svg}
\usepackage{bm}
\usepackage{braket}
\usepackage[title]{appendix}

\usepackage{xcolor}
\usepackage{mathtools}

\begin{document}

\preprint{APS/123-QED}

\title{Topological phase transitions via attosecond x-ray absorption spectroscopy}

\author{Juan F. P. Mosquera}
\affiliation{Departamento de Qu\'{i}mica, Universidad Aut\'{o}noma de Madrid, 28049 Madrid, Spain}
\author{Giovanni Cistaro}%
\affiliation{Departamento de Qu\'{i}mica, Universidad Aut\'{o}noma de Madrid, 28049 Madrid, Spain}

\author{Mikhail Malakhov}%
\affiliation{Departamento de Qu\'{i}mica, Universidad Aut\'{o}noma de Madrid, 28049 Madrid, Spain}
\affiliation{M.N. Mikheev Institute of Metal Physics of the Ural Branch of the Russian Academy of Sciences, S. Kovalevskaya str. 18, 620108 Yekaterinburg, Russia}

\author{Emilio Pisanty}
\affiliation{ICFO -- Institut de Ciencies Fotoniques, The Barcelona Institute of Science and Technology, 08860 Castelldefels (Barcelona), Spain}
\affiliation{Attosecond Quantum Physics Laboratory, King's College London, London WC2R 2LS, UK}

\author{Alexandre Dauphin}
\affiliation{ICFO -- Institut de Ciencies Fotoniques, The Barcelona Institute of Science and Technology, 08860 Castelldefels (Barcelona), Spain}

\author{Luis Plaja}
\affiliation{Grupo de Investigación en Aplicaciones del Láser y Fotónica, Departamento de Física Aplicada,
University of Salamanca, E-37008, Salamanca, Spain}

\author{Alexis Chac\'on}
\affiliation{Department of Physics and Center for Attosecond Science and Technology, POSTECH, 7 Pohang 37673, South Korea; Max Planck POSTECH/KOREA Research Initiative, Pohang, 37673, South Korea}

\author{Maciej Lewenstein}
\affiliation{ICFO -- Institut de Ciencies Fotoniques, The Barcelona Institute of Science and Technology, 08860 Castelldefels (Barcelona), Spain}
\affiliation{ICREA, Pg. Llu\'is Companys 23, 08010 Barcelona, Spain}

\author{Antonio Pic\'on}
\email{antonio.picon@uam.es, corresponding author}
\affiliation{Departamento de Qu\'{i}mica, Universidad Aut\'{o}noma de Madrid, 28049 Madrid, Spain}
\affiliation{Condensed Matter Physics Center (IFIMAC), Universidad Autónoma de Madrid, 28049, Madrid, Spain.}

\date{\today}

\begin{abstract}
We present a numerical experiment that demonstrates the possibility to capture topological phase transitions via an x-ray absorption spectroscopy scheme. We consider a Chern insulator whose topological phase is tuned via a second-order hopping. We perform time-dynamics simulations of the out-of-equilibrium laser-driven electron motion that enables us to model a realistic attosecond spectroscopy scheme. In particular, we use an ultrafast scheme with a circularly polarized IR pump pulse and an attosecond x-ray probe pulse. A laser-induced dichroism-type spectrum shows a clear signature of the topological phase transition. We are able to connect these signatures with the Berry structure of the system. This work extend the applications of attosecond absorption spectroscopy to systems presenting a non-trivial topological phase.
\end{abstract}

\maketitle


\section{\label{sec:intro}Introduction}

Topological states of matter permit the possibility to create materials that are insulators within the bulk, but they hold conducting states on the surface, known as edge states. Edge states are symmetry protected by the intrinsic properties of the bulk, providing a new landscape for developing unique optoelectronics applications with no precedence. In the last years, it has been enormous progress in the development of these modern materials, such as topological insulators \cite{Hasan2010,Bansil2016}. Interestingly, with the advent of ultrashort intense lasers, some materials that are topologically trivial insulators in equilibrium can be driven out of equilibrium into topologically non-trivial insulators \cite{Oka2009,Inoue2010,Sie2015,McIver2020,JimenezGalan2020, JimenezGalan2024,Tyulnev2024}. Those light-induced topological insulators, also known as Floquet topological insulators, only live during the laser pulse length, that is only around several femtoseconds (10$
^{-15}$ s). Generating topological states of matter is as important as characterizing them \cite{Ma2021}. In equilibrium, detecting if an insulator is topologically trivial or not trivial is possible by measuring the energy bands of the edge states via angle-resolved photoemission spectroscopy (ARPES) \cite{Bansil2016,Lv2019,Sobota2021}. Characterizing Floquet topological insulators is much more challenging, as we need an ultrafast probe in the femtosecond timescale in order to capture the ultrafast topological phases. One possibility is performing time-resolved ARPES and follow the edge-state energy bands in time \cite{Lv2019,Freericks2009,Sentef2015} or spin-resolved ARPES \cite{Sobota2021,Schuler2020,Schuler2020B}. However, the use of photoelectron spectroscopy techniques in the ultrafast regime have the drawback that reducing the probe pulse length decreases the photoelectron energy resolution. Additionally, if the system under investigation is interacting with a moderate intense IR pulse that drives the system out of equilibrium, the photoelectron spectrum becomes hard to interpret due to the laser effects on the ejected electron \cite{Cavalieri2007,Siek2017,Ossiander2018}. Another possibility is the use of high-order harmonic generation (HHG) \cite{Ghimire2019}. HHG is an extreme non-linear process in which several IR photons of the driving laser are absorbed and emitted by the material in the form of a high-frequency photon. Recent theoretical developments have shown that HHG is sensitive to the Berry curvature of the system \cite{Liu2017,Luu2018}, and the measurement of topological phases has been demonstrated \cite{Bauer2018,JimenezGalan2020,Silva2019,Chacon2020,Bai2020,Schmid2021,Baykusheva2021,Heide2022}. Despite the enormous progress in the last years, resolving ultrafast topological phases in periodic systems are far from being resolved.

Here, we present a complementary scheme to ARPES that enables us to directly measure the topological phase by using x-ray attosecond absorption spectroscopy. This scheme relies on a dispersive absorption measurement, which it is suitable for ultrashort probe pulses as the energy resolution is not diminished by the probe pulse length \cite{Leone2016}. Attosecond absorption spectroscopy has been successfully applied in different bulk and thin materials, from insulators to semimetals, to investigate carrier dynamics, phononic effects, and excitonic interactions \cite{Schultze2014,Lucchini2016,Zuerch2017,Moulet2017,Schlaepfer2018,Volkov2019,Lucchini2021,Buades2021}. In this work, we extend the application of attosecond absorption spectroscopy to probing topological phase transitions.


\begin{figure*}
\centering\includegraphics[scale=0.4]{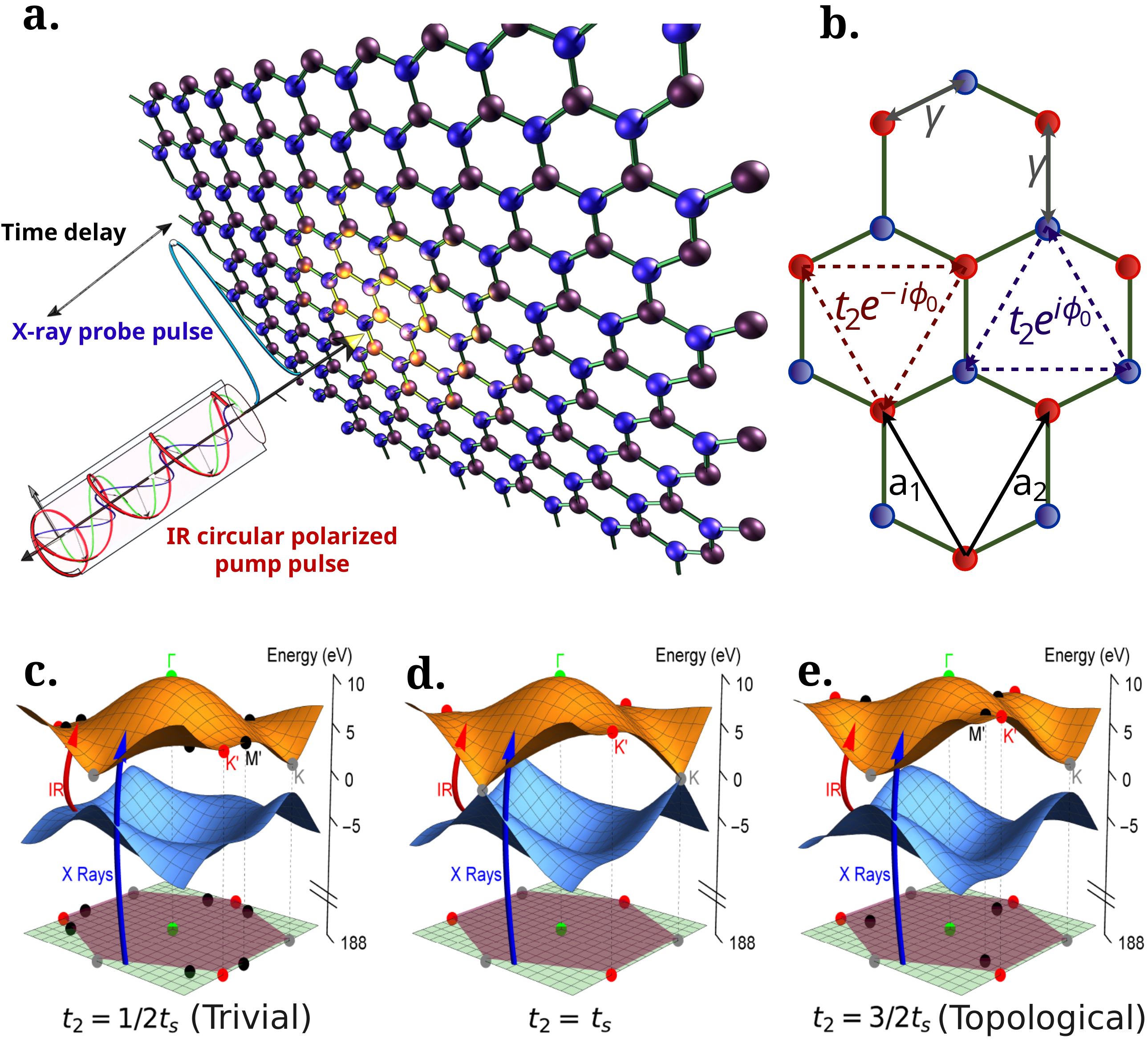}
\caption{Ultrafast x-ray scheme and system under investigation. (a) Two ultrashort laser pulses, separated by a time delay, interact with a boron nitride monolayer. The pump pulse is in the range of mid IR and is circularly polarized, and it is intense in order to drive a strong intra-band current. The probe pulse is in the range of soft x-rays and is linearly polarized, and it excites transitions from the K-edge of boron. (b) Graphical representation of the first ($\gamma$) and second order hopping ($t_2e^{\pm i\phi_0}$) terms between different lattice sites. (c), (d) and (e) represent the energy dispersion of the system. The red and blue arrows represent the IR and x-ray transitions, respectively. The IR pulse couples the conduction and valence bands, while the x-ray pulse couples the core band (1s orbital of boron) with the bands around the Fermi level. The second-order hopping $t_2$ controls the topological phase, which depends on the parameter $t_s=\Delta/6\sqrt{3}$, being $\Delta$ the gap of the system without second-order hopping. (b) and (d) are insulators with a trivial and non-trivial topology, respectively, and their direct bandgap is the same. In (d) the conduction and valence bands join at the K point. 
}
\label{fig:scheme}
\end{figure*}

The absorption of an attosecond pulse occurs in a very short timescale, in which the dynamics are dominated by the electron response. The induced electron motion is therefore much faster than the nuclei motion, and we may consider neglecting the coupling with phonons at such timescale. We can then conceive experiments that take advantage of the coherent electron dynamics \cite{Mandal2021,Xue2022,Moitra2023}. In a previous work \cite{Cistaro2021}, we demonstrated that the attosecond absorption is very sensitive to Van Hove singularities, and it is possible to extract information around these points, not only of the energy-dispersion structure, but also of the Berry structure. By performing numerical simulations with a recently developed approach to describe the electron dynamics out of equilibrium, the EDUS code \cite{Cistaro2023}, we show here that attosecond absorption spectroscopy can indeed capture topological phase transitions. This work is based on a Chern insulator described by a Haldane Hamiltonian \cite{Haldane1988}, in which we control the topological phase by changing a second-order hopping. We correlate the features of the absorption to the laser-driven coherent electron dynamics by using a semiclassical approach \cite{Cistaro2021,Picon2019,Dong2022}. This enables us to get a further insight of the effects of the Berry structure on the absorption spectrum around van Hove singularities. Our numerical and theoretical study opens the door to further investigations on relevant systems for optoelectronics applications, such as topological or Floquet insulators.

\section{\label{sec:theory} Ultrafast laser-induced x-ray dichroism}

We aim at studying the signatures of electron dynamics on the absorption of attosecond x-ray pulses for different topological phases. We start with a Haldane Hamiltonian \cite{Haldane1988} for a tight-binding model of boron-nitride monolayer (hBN), and by adding a second-order hopping that breaks time-reversal symmetry (TRS) we manipulate the topological phase. The system interacts with a linearly-polarized x-ray pulse and a circularly-polarized intense IR pulse. The Hamiltonian is expressed as $H(t)=H_0 + V(t)$, where $H_0$ is the Haldane Hamiltonian and $V(t)$ is the light-matter interaction coupling. More details are given in appendix \ref{sec:hamiltonian}. From previous studies \cite{Picon2019,Cistaro2021,Dong2022}, we know that the lineshape of an attosecond absorption spectrum is modified by the laser-driven electron dynamics, particularly at the energies corresponding with van Hove singularities. Importantly, the energy and Berry structure of those points are sensitive to the topological phase. This inspires us to use a laser-induced x-ray dichroism scheme as the one illustrated in Fig. \ref{fig:scheme}. The IR pulses induces strong intra-band currents, making electrons to follow circular trajectories. By changing the polarization handedness, and observing the difference in the absorption lineshape, i.e. a dichroism-type absorption spectrum, we infer information of the Berry structure (connected with the topological phase). This is because left-handed and right-handed laser-driven trajectories are not so different in energy, overall, but they are with respect to the Berry structure due to the second-order hopping, see the illustration in Fig. \ref{fig:scheme}(b), as we show in the following.

In the Haldane Hamiltonian, when the second-order hopping is $t_2= t_s=\Delta/6\sqrt{3}$, being $\Delta$ the gap of the system without second-order hopping, the conduction and valence bands join at the K points, as illustrated in Fig. \ref{fig:scheme}(d). For larger (smaller) hopping, the gap opens, and the system presents a non-trivial (trivial) topology. The x-ray pulse mainly promotes core electrons from the boron site to the conduction band, as the valence band is fully occupied before the arrival of the IR pulse. The attosecond x-ray pulse of 188 eV photon energy is linearly polarized, with a small intensity to ensure a first-order perturbation excitation. The polarization is out-of-plane in order to couple the 1s-2p transitions. The pulse length is 80-as and an intensity of 10$^{10}$ W/cm$^2$, short enough in order that the bandwidth, with a full-width half-maximum (FWHM) of 188 as, covers the energy bands around the Fermi level. Core electrons are further driven along the conduction band in the presence of the IR pulse before the core-hole relaxation that occurs approximately in 6.91 fs, translated to a decay width $\Gamma_{ch}=0.095$ eV. In this short timescale, the electron dynamics is coherent and the accumulated dynamical phase is imprinted in the attosecond absorption spectrum \cite{Cistaro2021}. The IR pulse of 0.41 eV photon energy (3000-nm wavelength and period of T$\sim10$ fs) is circularly polarized, with an intensity of 10$^{11}$ W/cm$^2$, and a pulse duration of 7 cycles, with an envelope modeled by a $\sin^2$ profile. Note that the time delay between both pulses are taken with respect to the maxima of each pulse envelope. Therefore, a negative time delay means that the probe maximum arrives to the system before the pump maximum. 

\begin{figure}
\centering\includegraphics[scale=0.99]{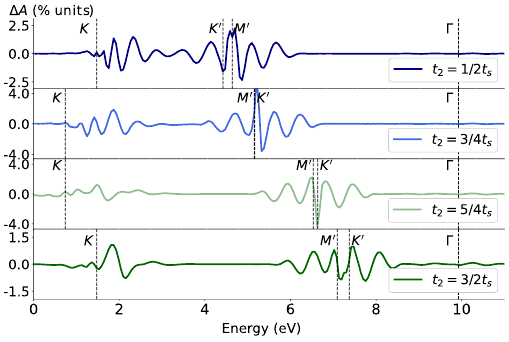}
\caption{Laser-induced x-ray absorption dichroism for different topological phases (a) $t_2= \frac{1}{2}t_s$, (b) $t_2= \frac{3}{4}t_s$, (c) $t_2= \frac{5}{4}t_s$, and (d) $t_2= \frac{3}{2}t_s$. The changes are localized around van Hove singularities (K, K', M', and $\Gamma$ points). $\Delta A$ expressed in percentage indicates the difference of absorption intensity normalized with respect to the maximum absorption calculated without the presence of the IR pulse.
}
\label{fig:dichroism}
\end{figure}

\begin{figure}
\centering\includegraphics[scale=0.99]{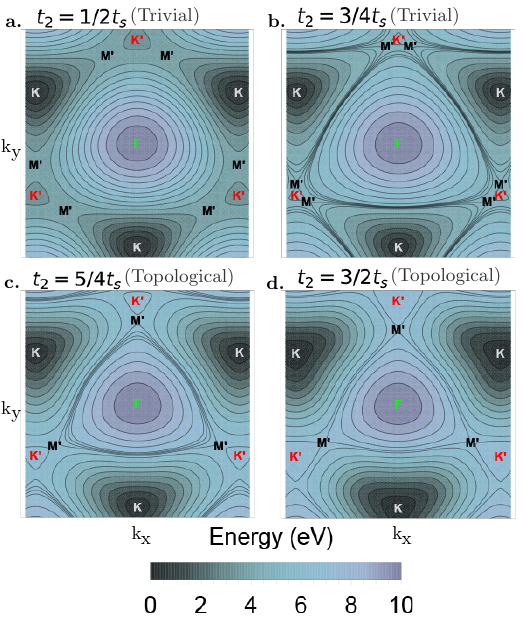}
\caption{Energy dispersion of the conduction band for (a) $t_2= \frac{1}{2}t_s$, (b) $t_2= \frac{3}{4}t_s$, (c) $t_2= \frac{5}{4}t_s$, and (d) $t_2= \frac{3}{2}t_s$. M' is a saddle point and changes with $t_2$. Note that the energy color scale takes as a 0-eV reference the K point, in which the direct gap is located for each case.
}
\label{fig:kpoints}
\end{figure}

The out-of-equilibrium dynamics induced by the lasers is calculated using our EDUS code \cite{Cistaro2023}. From the calculated electron dynamics, we obtain the polarization of the system in time, and from this one we obtain the absorption of the system. The laser-induced x-ray absorption dichroism for different topological phases is shown in Fig. \ref{fig:dichroism}. This has been calculated, for a fixed time delay ($\tau_d=$ 0.0 fs) between the IR and the x-ray pulse, by computing the absorption spectrum when the IR pulse is left-handed and right-handed circularly polarized and then taking the difference between them. If the system has no second-order hopping, i.e. $t_2=0$, then there is no dichroism signal. When $t_2\neq0$, the absorption dichroism is localized around the van Hove singularities. We clearly observe that the energies corresponding to the K, K', and M' points are those with stronger signal. The M' point is located in different points of the reciprocal space depending on the topological phase, see Fig. \ref{fig:kpoints}. The M' point coincides with the M point when there is no second-order hopping and represents a saddle point. Note that the M' points are between the K' and K points in the trivial phase, and between K and $\Gamma$ points in the topological phase.

When the system is a trivial or non-trivial insulator, the lineshape of the absorption dichroism around the K point is quite different, having a stronger multipeak structure when the system is a trivial insulator. The signal around M' and K' overlap, but in general do not show such strong variation when the topology of the system changes. The energy positions of the van Hove singularities changes due to topology, this is expected because the energy dispersion depends on $t_2$, see Fig. \ref{fig:kpoints}. Interestingly, in Fig. \ref{fig:dichroism} we show two different examples, top and bottom panel, in which the energy gaps are the same, but still the calculated lineshapes are very different when the system has a trivial or a topological phase. Note that the signals are around 1-4 \% units of the maximum calculated absorption without the presence of the IR pulse, so these changes are notable. We have also explored different IR intensities. As higher the intensity, the signal is higher and the multipeak structure extends in energy (showing more modulations), but in general the same behavior is observed around the K points during a topological phase transition. 

\begin{figure*}
\centering \includegraphics[scale=0.25]
{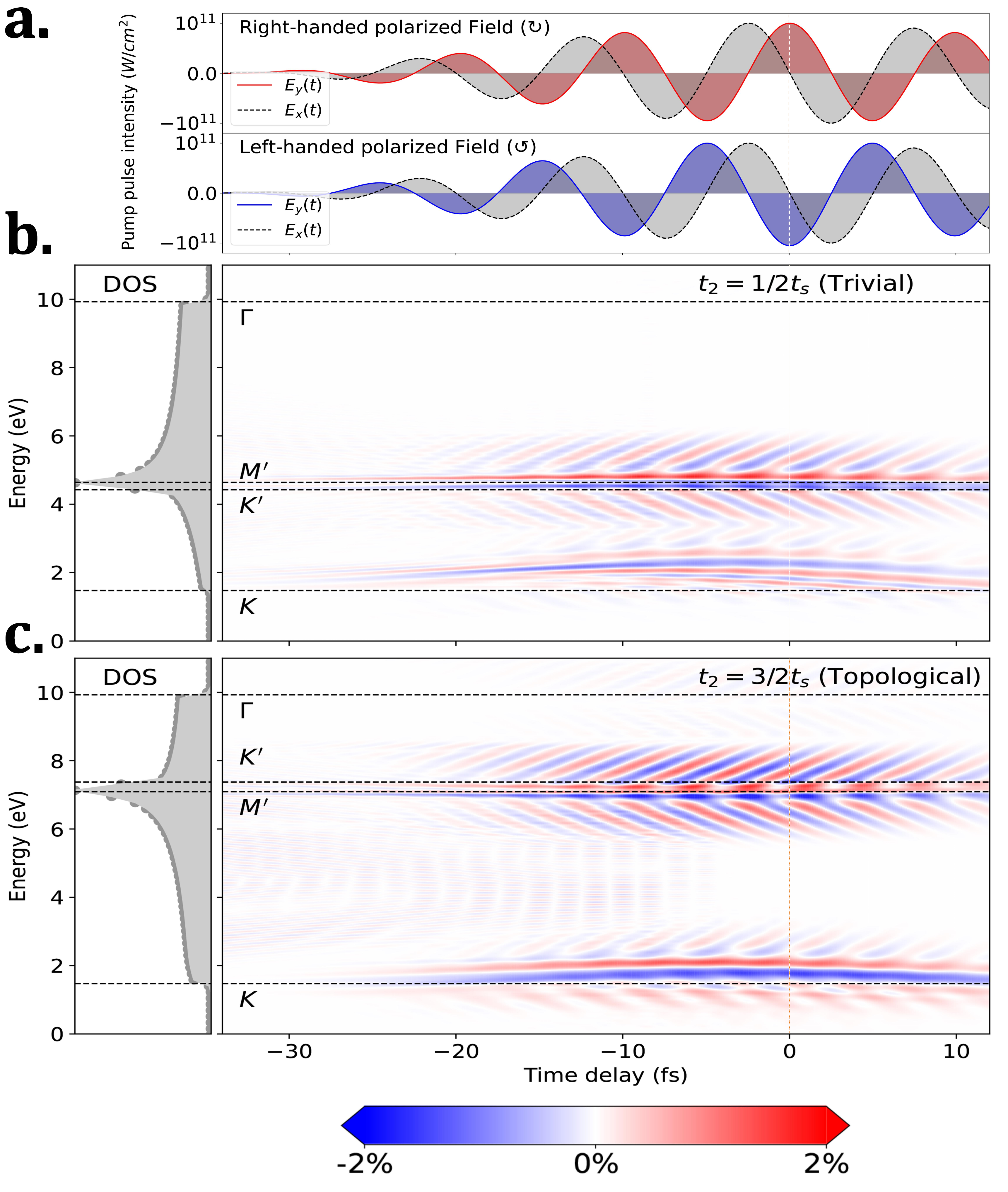}
\caption{Laser-induced x-ray absorption dichroism for different pump-probe time delays. (a) Scheme of the IR pumb pulse for right- and left-handed polarized fields. Ultrafast laser-induced x-ray dichroism for (b) a trivial, $t_2= \frac{1}{2}t_s$, and (c) a topological, $t_2= \frac{3}{2}t_s$, phase. The density of states (DOS) is represented on the side. 
}
\label{fig:atas}
\end{figure*}

The laser-induced dichroism shown in Fig. \ref{fig:dichroism} is for a fixed time delay between the two laser pulses. We also explore different time delays $\tau$, on which a negative value means that the maximum of the probe pulse occurs before the maximum of IR pump pulse. Thus, a zero time delay $\tau=0$ is understood as the maxima of both pulses overlap. The ultrafast laser-induced dichroism at different time delays is shown in Fig. \ref{fig:atas} for trivial and topological phases. The time-dependent changes around the M' and K' points are located at different energies when the system is trivial ($t_2=t_s/2$) or topological ($t_2=3t_s/2$). This is because of the difference in the energy dispersion. However, the time-dependent changes are quite similar. The time-dependent changes around the K point are quite different for the two different phases though, the topological phase presents two (positive, red, and negative, blue) clear peaks, while the trivial phase presents a higher modulation (multipeak structure). These changes are also influenced by the Stark shift induced by the laser pulse and depends on the IR intensity.

In conclusion, the used ultrafast scheme clearly shows a signature when the topological phase changes. In the next section, we analyze the absorption dichroism using a semiclassical model that enables us to understand the laser-induced lineshapes and correlate them to the Berry structure.

\section{\label{sec:numerics}Correlating the Berry structure with the laser-induced dichroism}

In our previous work \cite{Cistaro2021}, we developed a semiclassical approach to calculate the attosecond absorption spectrum. In this approach, the polarization of the system is calculated by the sum of semiclassical electron trajectories driven by the IR vector potential. The approach assumes that an electron starts in a particular ${\bf k}$ point of the conduction band, i.e. the excitation by the attosecond pulse is sudden, and the quasi-momentum evolves as ${\bf K}={\bf k}-{\bf A}(t)$, where ${\bf A}(t)$ is the vector potential of the IR pulse. Hence, the intra-band current is described by the evolution of the quasi momentum, and the contribution of the valence band is neglected. Along the semiclassical trajectory, the electron acquires a phase given by 
\begin{eqnarray}
\Delta\varphi(t,t_0,{\bf K}) = i  \int_{t_0}^t dt' [  (\epsilon_{c}({\bf K} + {\bf A}(t')) - \epsilon_{c}({\bf K}))] \nonumber \hspace{.5cm}\\
+ i  \int_{t_0}^t dt'  [\Phi_{S} ({\bf K} + {\bf A}(t'),t') ] \hspace{.5cm}
\label{eq:phase}
\end{eqnarray} 
where the dynamical, action Berry phase $\Phi_{S}$ is $
\Phi_{S} ({\bf k},t)=  \frac{1}{2} \pmb{\varepsilon}_{IR}(t) \cdot  \partial_{\bf k}  \phi({\mathbf{k}})  [ \cos \theta({\mathbf{k}}) + 1] 
$, being $\epsilon_{c}({\bf k})$ the energy of the conduction band, and the angles $\theta({\bf k})$ and $\phi({\bf k})$ are related to the Berry structure of the system, see appendix \ref{sec:hamiltonian}. Note that the coherent phase mainly depends on the conduction band and not at all on the core-hole band $\epsilon_{ch}({\bf k} )$. This is due to the fact that the core orbitals are well localized and, therefore, the electronic structure is not ${\bf k}$ dependent.

\begin{figure*}
\centering \includegraphics[scale=0.40]{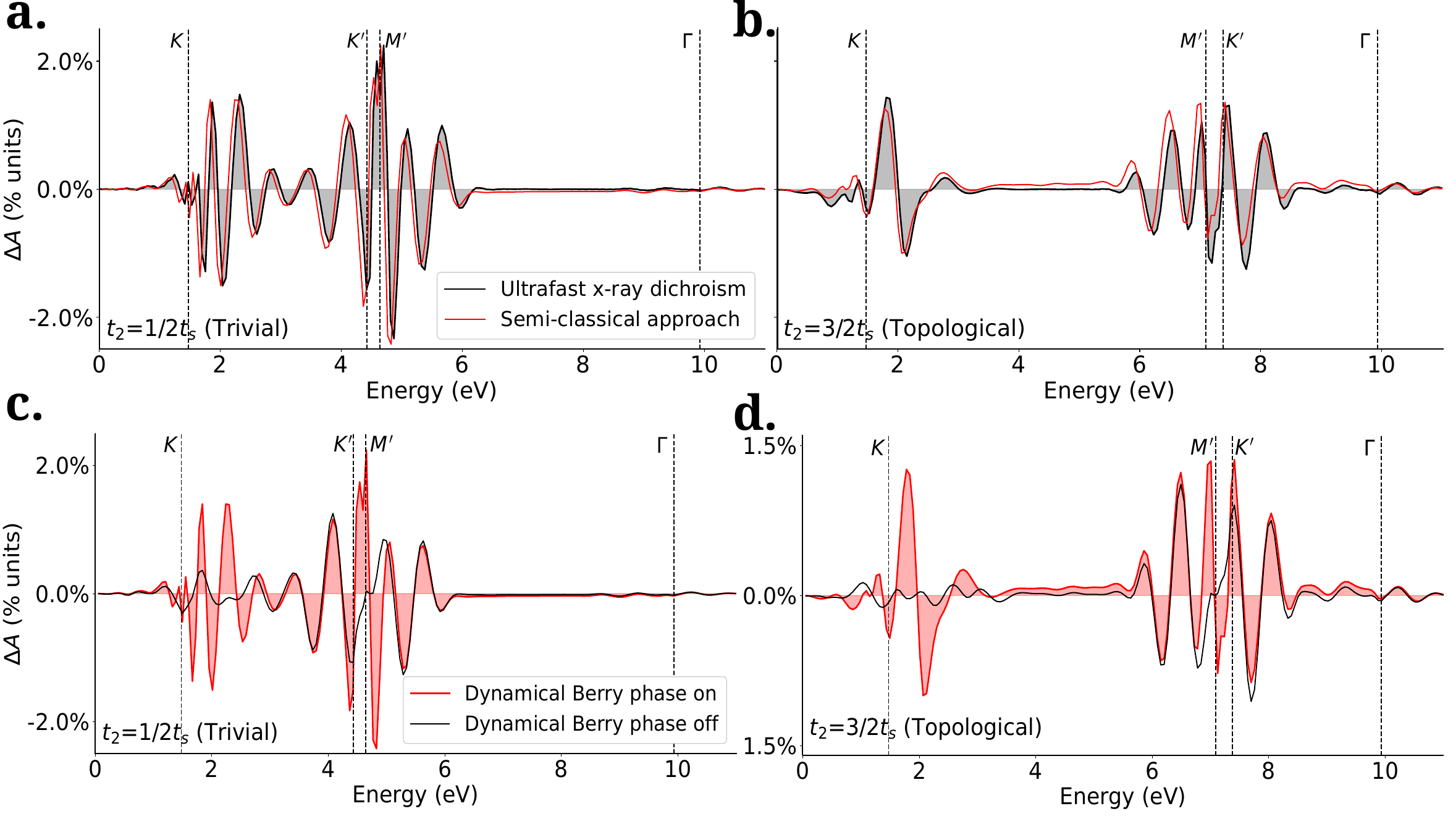}
\caption{Semiclassical approach for the laser-induced x-ray dichroism. (a),(b) Comparison of the laser-induced x-ray dichroism between first-principle and semiclassical calculations for trivial $t_2= \frac{1}{2}t_s$ and topological $t_2= \frac{3}{2}t_s$ phases. (c),(d) The laser-induced x-ray dichroism computed by the semiclassical approach with and without considering the Berry structure for trivial and topological phases. The red (black) line is when energy and Berry phase are included (when energy is only included) in the accumulated dynamical phase, see Eq.~(\ref{eq:phase}).
}
\label{fig:semiclassical}
\end{figure*}

Once we calculate the phase for different trajectories at different points of the reciprocal space, then we calculate the polarization of the system as 
\begin{align}
\label{eq:polarization}
{\bf P}(t) \! \propto \! \sum_{{\bf K}} \, &\left[ \frac{ \left|\pmb{\xi}_{20}(\textbf{K}+\textbf{A}(t))\right| } { \left|\pmb{\xi}_{20}(\textbf{K}+\textbf{A}(t_0)) \right| }\right.\nonumber \\ &\left. e^{-i(\epsilon_{ch}({\bf K} ) - \epsilon_{c}({\bf K}) - i \Gamma_{ch}/2)(t-t_0) + \Delta\varphi(t,t_0,{\bf K})}
+ c.c. \right], \nonumber \\ 
\end{align}
where we sum over all the ${\bf K}$ points, and $\pmb{\xi}_{20}$ is the inter-band Berry connection matrix that couples the core and conduction band, see more details in appendix \ref{sec:berry_connection} and \ref{sec:semiclassical}. The previous equation can be easily understood as the sum of all the oscillating dipole terms in the reciprocal space, which deviates by a phase $\Delta\varphi$ that arises from the light-induced electron motion during the core-hole decay, represented by the width $\Gamma_{ch}$. 

Once we have the semiclassical polarization, Eq.~(\ref{eq:polarization}), we obtain from it the attosecond absorption spectrum \cite{Cistaro2023}, both when the IR laser pulse is left-handed and right-handed circularly polarized, and take the difference in order to obtain the laser-induced absorption dichroism, see  Fig.\ \ref{fig:semiclassical} for the case of a trivial and topological phase. The agreement with the first-principle calculations obtained with EDUS, see Figs.\ \ref{fig:semiclassical}a,b, is excellent. This simple semiclassical model is then enough to describe the observed features of the dichroism spectrum. The advantage of the semiclassical approach is that it enables us to distinguish the effects of the energy and the Berry structure of the system. We compute the laser-induced absorption dichroism, but now by not including the dynamical Berry phase in the accumulated phase of Eq.~(\ref{eq:phase}), see the black line in Figs.\ \ref{fig:semiclassical}b,c. 

The latter calculations show that the dichroism features around the K point are mainly because of the Berry structure of the material.  The accumulated energy phase is very similar for left-handed and right-handed circular polarization. This phase induces an energy shift towards higher energies. The accumulated action Berry phase changes sign due to the handedness of the polarization. The part of the dynamical Berry phase that depends on the angle $\theta$ changes the sign for the topological case compared with the trivial case. This change of sign has consequences when is added to the other phase terms. In the topological phase, the dynamical Berry phase is comparable to the energy phase and results in the two well-separated (positive and negative) peaks, Fig.\ \ref{fig:semiclassical}b, while for the trivial phase, the Berry phase is weak and results in the multipeak structure, Fig.\ \ref{fig:semiclassical}a. 

We can get a different point of view by relating the dynamical Berry phase with the Berry curvature. If we assume a closed electron trajectory, i.e. the electron is in the same position in the reciprocal space at the initial time $t_i$ and at the final time $t_f$, then the accumulated dynamical phase is  $\int_{t_i}^{t_f} dt'  [\Phi_{S} ({\bf K} + {\bf A}(t'),t') ] = \int_{\Delta S} d{\bf S} \cdot \boldsymbol{\Omega} ({\bf K})$, where $\Delta S$ is the area enclosed by the trajectory, ${\bf S}$ is a normal vector to the area $\Delta S$, and $\boldsymbol{\Omega} ({\bf K})$ is the Berry curvature of the conduction band. If the enclosed trajectory is small enough that the Berry curvature is constant within the area, then one expects that the accumulated phase is proportional to the Berry curvature. The Berry curvature changes sign around the K point when there is a topological phase transition. 

Around the K' and M' points, we observe that both the energy and the dynamical Berry phase play an important role. The laser-driven electron trajectories perceive then a different energy-dispersion landscape depending on the handedness of the polarization, as it takes place around the M' point, see the energy dispersion represented in Fig.~\ref{fig:kpoints}. Hence, the energy dispersion has an important effect on the signal around the K' and M' energy points. 

In conclusion, the semiclassical approach enables us to correlate the changes in the absorption spectrum with the laser-driven dynamics. This correlation allows us to understand the observed features around the different van Hove singularities as a function of the topological phase.

\section{\label{sec:examples}Saddle Point Approximation}

To get a clear physical insight of the absorption spectrum around the K point (van Hove singularity with stronger differences under a phase transition), we have studied the nature of different trajectories driven by the IR pulse in the framework of the semiclassical approach, knowing that the polarization of the system would be given by $P(\omega) = \int_{-\infty}^{\infty} dt e^{-i\omega t} P(t)$.
Due to the nature of our specific pump-pulse scheme, it is possible to assume that the exponential action $e^{-iS} \equiv e^{-iS(\textbf{k},t,t_0)}$ (defined in detail and its connection to the polarization in the appendix~\ref{sec:semiclassical}) oscillates rapidly as a function of the crystal momentum $\textbf{k}$. Therefore, one can perform a saddle-point approximation (SPA) \cite{Becker2006,Nayak2019,Lewenstein2020,Milosevic2020}, to find the trajectories in the reciprocal space that satisfy with the conditions

\begin{widetext}
\begin{align}
    \Delta \textbf{R} = {\bf r}_d ({\bf K},t,t_0) - \pmb{\alpha}({\bf K}) + \pmb{\alpha}^\parallel (\textbf{K}-{\bf A}(t)+{\bf A}(t_0))= 0,
    \label{eq:saddle_point_trajectory}\\
    \epsilon_c({\bf K}) - \epsilon_{ch}({\bf K}) + \pmb{\varepsilon}_{IR} (t)\cdot \left[ {\bf r}_d ({\bf K},t,t_0) +  \pmb{\alpha}^\parallel (\textbf{K}-{\bf A}(t)+{\bf A}(t_0)) \right] = \omega,
\label{eq:saddle_point_energy}
\end{align}
\end{widetext}
which are similar to the ones obtained in the study of the recollision model for HHG in solids \cite{Gaarde2020,Yue2022}. In this approximation, we are neglecting the effects of the core-hole decay $\Gamma_{ch}$. Also, it is more convenient to rewrite the dynamical action Berry phase Eq.~(\ref{eq:phase}) explicitly as a function of the inter- and intra-band Berry connection matrix elements to study the contribution of each of these terms, see more details in Appendix \ref{sec:SPA}. These equations are easy to intepret. On one hand, the first equation is interpreted as the distance between the electron-hole pair, $\Delta \textbf{R}$. When this distance is zero, then the electron and hole may recombine. The trajectories satisfying this condition are the most relevant in the action then. Interestingly, the term ${\bf r}_d (\textbf{K},t,t_0)$ contains the ballistic velocity, which just depends on the energy dispersion of the conduction band, but also the anomalous velocity, which depends on the Berry curvature of the conduction band. The Berry distances $\pmb{\alpha}({\bf k})$ depend purely of the intra- and inter-band Berry connections. These terms modify when the recombination may occur. This is an important term to describe the so-called imperfect recollisions in HHG \cite{Gaarde2020}. On the other hand, the second equation is interpreted as the energy during recombination. It will provide the energy $\omega$ of the absorption when the recombination occurs. 

\begin{figure}
\centering\includegraphics[scale=0.45]{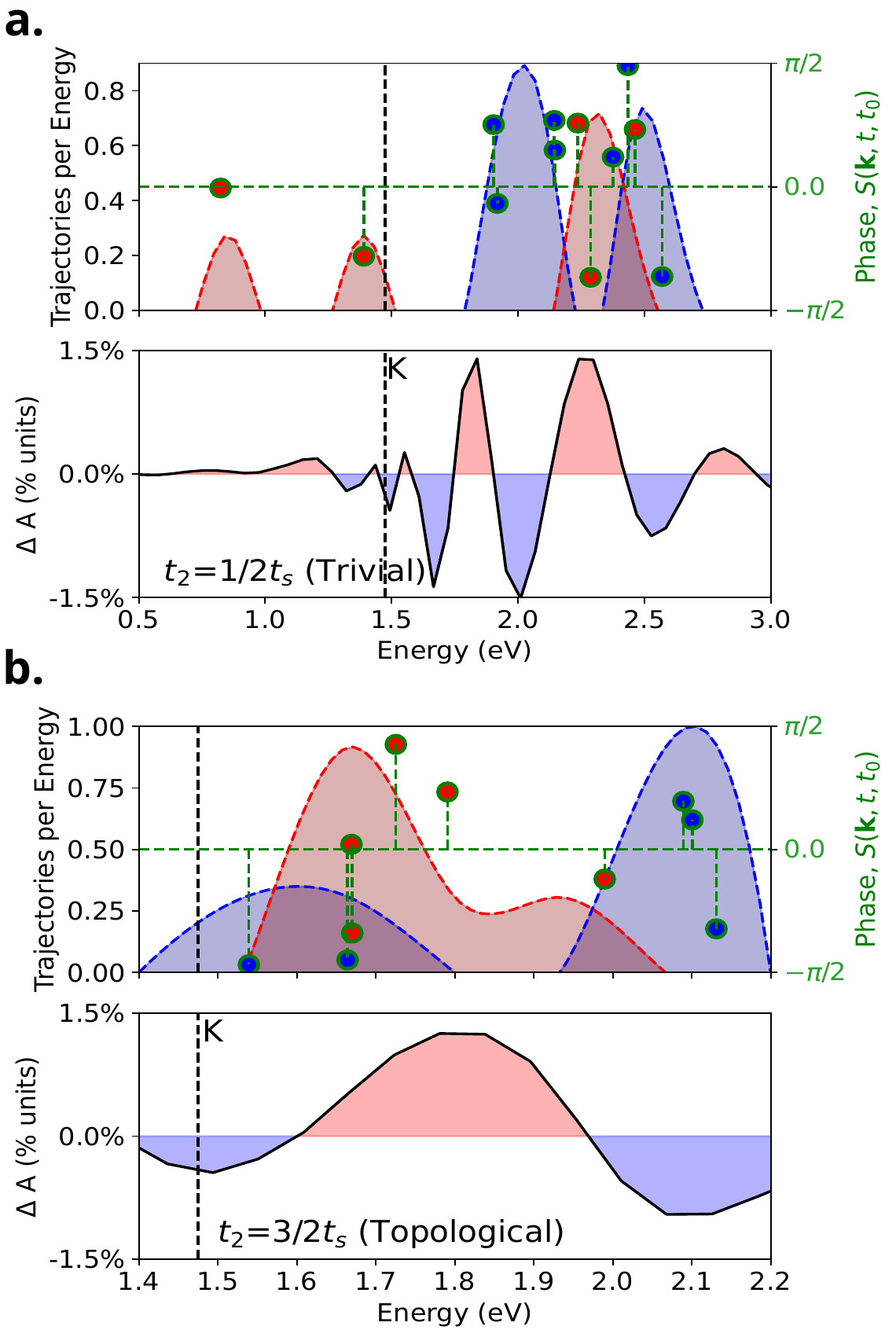}
\caption{In the top panels, the dots represent the trajectories that satisfy the SPA conditions for (a) a trivial $t_2= \frac{1}{2}t_s$ and (b) a topological $t_2= \frac{3}{2}t_s$ phase. The trajectories for right-handed (left-handed) circular polarization are highlighted by red (blue) dots.  In order to better visualize the number of trajectories per energy, we add a broadening to each one that sum up incoherently. We also represent the accumulated phase of each trajectory at their corresponding recolision time in green. In the bottom panels, the laser-induced x-ray dichroism spectrum around the neighborhood of K is shown for comparison.
}
\label{fig:Phases}
\end{figure}

To compute the time evolution of different trajectories at different initial points in the {\bf k}-space, we use a discretized {\bf k}-grid in the neighborhood of the K point, as the guess for the initial positions of the electron in the conduction band, and then evolve them using the quasimomentum $\textbf{K}$. Only those trajectories which recombine ($\Delta \textbf{R} =0$) at a certain time contribute to the polarization. Thus, the recollision times $t$ and the initial crystal momenta $\textbf{k}$ can be found using equation (\ref{eq:saddle_point_trajectory}). Among all possible trajectories, just the ones that recollide at a certain energy $\omega$, see Eq.~(\ref{eq:saddle_point_energy}), are shown in Fig.~\ref{fig:Phases}. 
The different dots represent the found trajectories at their corresponding recollision energy $\omega$, both for right-handed circular polarization (red dots) and left-handed circular polarization (blue dots).  In order to better visualize the number of trajectories per energy, we add a broadening to each one that sum up incoherently. In the top of that, we have added, in green, the value of the corresponding action, $S(\textbf{k},t,t_0)$ (accumulated imaginary phase) at the given recollision time of each trajectory. 

For both phases, trivial and topological, the contribution of the Berry curvature in the integral of the displacement ${\bf r}_d ({\bf K},t,t_0)$ is small. Also, the contribution of the intra-band Berry connections in $\pmb{\alpha}(\textbf{K})$ (or $\pmb{\alpha}^\parallel(\textbf{K})$) is larger than the inter-band Berry connections for computing $\Delta \textbf{R}$ in the trajectories that recollide. Therefore, the trajectories that satisfy the first condition of the SPA are mostly described by the intra-band Berry connection $\pmb{\xi}_{22}(\textbf{K})$ and the dispersion velocity $\textbf{v}(\textbf{K}) =  \nabla_{\bf K} \Delta E(\textbf{K}) = \nabla_{\bf K} \epsilon_c(\textbf{K})$. The latter is responsible for most of the effects, i.e. the highest contribution is given by how electrons move along the dispersion energy rather than by how the bands are coupled by the infrared pulse. Thus, the energies $\omega$ in Eq.~(\ref{eq:saddle_point_energy}) are described by the energy $\epsilon_c({\bf k})$ and $\pmb{\alpha}(\textbf{K})$ ($\pmb{\alpha}(\textbf{K}) \approx \pmb{\xi}_{22} (\textbf{K}) $), the latter modulated by the electric field at the recollision time.


For the topological phase in the neighborhood of K, the trajectories whose energies are lower than $1.7$ eV (their initial $\textbf{k}$ are not further $0.1$ a.u.\ from K), their recollision times are larger than $11$ fs., then the stronger contribution to the energy $\omega$ is just given by $\epsilon_c({\bf k})$. On the other hand, for those trajectories whose energy is larger than $1.7$ eV, both energy and intra-band Berry connection contribute, the latter modulated by the amplitude of the electric field, which groups the trajectories together towards smaller energies for the left-handed polarized IR pulse and towards larger energies for the right-handed case. This may qualitatively explain the two-single peak structure of the light-induced dichroism spectrum just in terms of the number of trajectories, see Fig. \ref{fig:Phases}a.

For the trivial phase in the region of energies around K, the contribution arising from the intra-band Berry connections is more notorious (both for those closer and further from the K point), and its effect, again modulated by the amplitude of the electric field at the different recollision times, results in trajectories recolliding at different energies, but being overlapped, see the trajectories in Fig.~\ref{fig:Phases}a. This may qualitatively related to the multi-peak structure in the light-induced dichroism spectrum, see Fig. \ref{fig:Phases}b.

We note that the accumulated phase of a single SPA trajectory may play an important role in the absorption spectrum. The analysis above is mainly restricted to the number of trajectories satisfying the SPA conditions, but each trajectory indeed accumulate a phase given by the function of the action. Because the absorption is sensitive to the imaginary part of the polarization \cite{Cistaro2021}, those trajectories approaching to zero accumulated phase may have a small contribution in the absorption spectrum. The accumulated phase is indicated in Fig. \ref{fig:Phases} for each trajectory.

In conclusion, the most important effect in the dynamical phase is coming from the contribution of the energy dispersion, but this is mainly modified by the intra-band Berry connection, which depends on the amplitude and polarization of the electric field at different recollision times. This latter effect is different for the case of trivial and topological phase.


\section{\label{sec:conclusions}Conclusions}

In this work we perform a numerical experiment and develop a semiclassical theory to investigate the effects of the topological phase on the absorption of an x-ray attosecond pulse. In particular, we propose an ultrafast laser-induced x-ray dichroism scheme to study a Chern insulator. The ultrafast scheme consists of an x-ray attosecond pulse, which excites core electrons into the conduction band, and a secondary IR pulse, which produces a strong intra-band motion. The IR pulse is left-handed and right-handed circularly polarized, and by calculating the x-ray absorption spectrum for the two IR polarizations, and then taking the difference between them, we obtain the absorption dichroism. We show that this scheme is very sensitive to the change of topological phase. The absorption dichroism features are localized around the energies corresponding to van Hove singularities. We develop a semiclassical theory to understand these features, correlate them with the electron dynamics around the van Hove singularities, and associate them with an accumulated dynamical phase that depends on the topology of the system. We show that the main changes due to the topological phase are because the Berry structure of the system. To further simplify the semiclassical model, we use the saddle-point approximation. This enables us to qualitatively study the main changes in the action and identify the intra-band Berry connection of the conduction band as the relevant contribution to the topological phase.


This work opens the door to further investigations in capturing coherent electron dynamics via attosecond absorption spectroscopy and infer relevant properties such as the Berry structure or the topological phase. Future investigations may involve the study of relevant systems for optoelectronic applications, such as Floquet and topological insulators \cite{Lein2022}.




{
\begin{appendices}

\section{Haldane hamiltonian} \label{sec:hamiltonian}

We consider the protypical Haldane Hamiltonian \cite{Haldane1988} in order to describe different topological phases. This Hamiltonian breaks time-reversal symmetry via a second-order hopping, typically used on a 2D hexagonal system. Here we consider a tight-binding hBN model \cite{Rivas2013}, with lattice vectors ${\bf a}_1=\frac{a}{2} (\sqrt{3} , -1)$ and ${\bf a}_2=\frac{a}{2} (\sqrt{3} , 1)$, being $a = 2.5$ \AA \, the lattice constant. For periodic boundary conditions, the Haldane Hamiltonian in the reciprocal space is given by
\begin{equation}
    \hat{H}_s(\textbf{k}) = B_0(\textbf{k}) \hat{\sigma}_0 + \sum_{i=1}^{3} B_i (\textbf{k}) \hat{\sigma}_i,
    \label{eq:hamiltonian_haldane_pauli}
\end{equation}
where we have used the fact that any Hermitian complex $2 \times 2$ matrix can be expressed in terms of the identity $\hat\sigma_0$ and the Pauli matrices $\hat\sigma_i$. 
The set of $B_0(\textbf{k})$ and $B_i(\textbf{k})$ is read as
\begin{align}
    B_{0}(\textbf{k}) & = 2t_2\cos{\left(\phi_0 \right)}\sum_{i}{\cos{\left( \textbf{k}\cdot \textbf{b}_i\right)}}, \\
    B_{1}(\textbf{k}) & = \gamma \sum_{i}{ \cos{\left( \textbf{k}\cdot \pmb{\delta}_i\right)} }, \\
    B_{2}(\textbf{k}) & = \gamma \sum_{i}{ \sin{\left( \textbf{k}\cdot \pmb{\delta}_i\right)} }, \\
    B_{3}(\textbf{k}) & = \frac{\Delta}{2} - 2t_2\sin{\left(\phi_0 \right)}\sum_{i}{\sin{\left( \textbf{k}\cdot \textbf{b}_i\right)}}.
\end{align}
being $\textbf{b}_i$ the second neighbors atoms and $\pmb{\delta}_i$ are the nearest neighbors, which can be writen as
\begin{equation}
    \begin{matrix*}[l]
        \pmb{\delta}_3 = \frac{a}{\sqrt{3}} (1,0), & \pmb{\delta}_2 = \pmb{\delta}_3 - \textbf{a}_1, & \pmb{\delta}_1 = \pmb{\delta}_3 - \textbf{a}_2,
    \end{matrix*}
\end{equation}
and 
\begin{equation}
    \begin{matrix*}[l]
        \textbf{b}_1 = -\textbf{a}_1, &  \textbf{b}_1 = \textbf{a}_2, &  \textbf{b}_1 = \textbf{a}_1 - \textbf{a}_2.
    \end{matrix*}
\end{equation}
The energy gap is $\Delta=5.9$ eV, and $\gamma$ and $t_2 e^{i\phi_0}$ are the first and second order hopping, respectively, given by $\gamma = 3.16$ eV and $\phi_0 = \pi/2$.

The eigenvalues of $\hat{H}_s$ are given by $E_{\pm} (\textbf{k}) = B_0(\textbf{k}) \pm \sqrt{B_1(\textbf{k})^2 + B_2(\textbf{k})^2 + B_3(\textbf{k})^2}$. Hence, the conduction energy is given by $\epsilon_{c}(\textbf{k}) = E_{+} (\textbf{k})$ and the valence energy by $\epsilon_{v}(\textbf{k}) = E_{-} (\textbf{k})$. The eigenvectors can be written, in a suitable gauge for the rest of the formulas, as 
\begin{align}
    \left|u_{-}(\textbf{k})\right> &=
    \begin{pmatrix}
        e^{-i\phi({\textbf{k})}/2}\sin{ \frac{\theta(\textbf{k})}{2} }\\[6pt]
       - e^{+i\phi({\textbf{k})}/2}\cos { \frac{\theta(\textbf{k})}{2} }
    \end{pmatrix}, \nonumber \\
    \left|u_{+}(\textbf{k})\right> &=
    \begin{pmatrix}
        e^{-i\phi({\textbf{k})}/2}\cos { \frac{\theta(\textbf{k})}{2} }\\[6pt]
        e^{+i\phi({\textbf{k})}/2}\sin{ \frac{\theta(\textbf{k})}{2} }
    \end{pmatrix},
\end{align}
where the angles $\theta(\textbf{k})$ and $\phi(\textbf{k})$ are the spherical coordinates angles between the vector formed by $(B_1,B_2,B_3)$ and are given by
\begin{align}
    \tan \phi(\textbf{k}) &= \frac{B_2 (\textbf{k})}{B_1(\textbf{k})},\\
    \cos \theta(\textbf{k}) &= \frac{B_3(\textbf{k})}{\sqrt{B_1^2(\textbf{k})+B_2^2(\textbf{k})+B_3^2(\textbf{k}) }}.
\end{align}
To include core orbitals to describe the x-ray interactions in our ultrafast laser-driven dichroism scheme, we have added the 1$s$ orbitals (as a flat band) of boron to the Haldane model, where we have assumed that the 1$s$ orbitals are well localized and there is no overlapping with the others. Thus, the x-ray photon has enough energy to promote an electron in the 1$s$ orbital (or core band) to the conduction/valence band. Finally, the extended Hamiltonian can be written as
\begin{equation}
    \hat{H}=
    \begin{pmatrix}
        E_{1s}-E_f & | & 0 & 0 \\[3pt]
        \hline
        0 & | & H_{11}(\textbf{k}) & H_{12}(\textbf{k}) \\[6pt]
        0 & | & H_{12}^*(\textbf{k}) & H_{22}(\textbf{k})
    \end{pmatrix},
    \label{eq:hamiltonian_haldane}
\end{equation} 
where the different matrix elements $H_{ij}(\textbf{k})$ are the ones given by the Haldane Hamiltonian in Eq.~(\ref{eq:hamiltonian_haldane_pauli}).

A summary of the used parameters are in Table~(\ref{table:parameters}).

\begin{table}[h!]
    \centering
    \begin{tabular}{||l c ||}
        \hline
        Parameter & Value \\
        \hline \hline
        Lattice constant, $a$ & 2.5 \AA \\ \hline
        Band gap, $\Delta$ & 5.9 eV \\ \hline
        First-order hopping, $\gamma$ & 3.16 eV \\ \hline
        Second-order hopping & \\
        \multicolumn{1}{||c}{$\phi_0$} & $\pi/2$ \\
        \multicolumn{1}{||c}{$t_s$} & $0.567728$ eV  \\ \hline
        Core-Hole decay, $\Gamma_{ch}$ & 0.00396892 au \\ \hline
        Core band energy & $-188.03074636$ eV \\ \hline
    \end{tabular}
    \caption{Used parameters in the Haldane model.}
    \label{table:parameters}
\end{table}

\section{Berry Connection} \label{sec:berry_connection}

It is useful to represent the Berry connection in a Wannier basis to avoid discontinuities presented in the eigenstate basis, i.e. the one that diagonalizes the unperturbed Hamiltonian in the reciprocal space. In this representation, the Berry connection acquires the very simple form
\begin{equation}
    \pmb{\xi}^{(w)}=
    \begin{pmatrix}
        0 & 0 & \textbf{r}_{1s,2p} \\
        0 & 0 & 0 \\
        \textbf{r}_{1s,2p} & 0 & 0
    \end{pmatrix},
\end{equation}
on which have used the symmetry and parity of the orbitals (1$s$ for the core band and 2$p$ for the valence and conduction bands) and the atomic gauge to simplify the expression presented above. We have defined $\textbf{r}_{1s,2p}$ as the Berry connection between the core orbital (1$s$) and the outer orbital that forms the chemical
bond (2$p$). Note that the coupling between the 1$s$ and 2$p$ orbitals
\begin{equation}
    \textbf{r}_{1s,2p}  = \int d^3 \textbf{r} \phi_{1s}^* (\textbf{r}) \textbf{r} \phi_{2p} (\textbf{r}),
\end{equation}
is related to the x-ray pulse. Thus, only the out-of-plane component (named from now $\textbf{r}_a$) will be non-zero and is equal to $0.041$ \AA.

For representing the Berry connection in the eigenstate basis, it is necessary to compute a transformation similar to 
\begin{equation}
    \pmb{\xi}(\textbf{k})=\hat{U}(\textbf{k})\pmb{\xi}^{(w)}(\textbf{k})\hat{U}^\dagger(\textbf{k})-i\left( \partial_{\textbf{k}} \hat{U}(\textbf{k}) \right)\hat{U}^\dagger(\textbf{k}),
\end{equation}
being $\hat{U}(\textbf{k})$ the unitary matrix formed by the eigenfunctions of the Haldane Hamiltonian spanned over the columns, see (Eq.~\ref{eq:hamiltonian_haldane}). In this case, this kind of transformation must be done to include a phase factor that preserves the gauge covariance \cite{Silva2019PRB}. Thus, the Berry connection in the eigenstate basis is then given by
\begin{widetext}

\begin{eqnarray*}
    \hspace{-.cm}\pmb{\xi}={\scriptscriptstyle\left(
    \begin{array}{ccc}
        0 & -\textbf{r}_a e^{ i \phi (\textbf{k})/2} \cos \frac{\theta (\textbf{k})}{2} & \textbf{r}_a e^{ i \phi (\textbf{k})/2} \sin \frac{\theta (\textbf{k})}{2} \\[3pt]
        -\textbf{r}_a e^{- i \phi (\textbf{k})/2} \cos \frac{\theta (\textbf{k})}{2} & -\frac{1}{2} \cos \theta (\textbf{k}) \partial_{\textbf{k}}\phi (\textbf{k}) & \frac{1}{2} \left(\sin \theta (\textbf{k}) \partial_{\textbf{k}}\phi (\textbf{k})-i \partial_{\textbf{k}}\theta (k)\right) \\[3pt]
        \textbf{r}_a e^{- i \phi (\textbf{k})/2} \sin \frac{\theta (\textbf{k})}{2} & \frac{1}{2} \left(\sin \theta (\textbf{k}) \partial_{\textbf{k}}\phi (\textbf{k})+i \partial_{\textbf{k}}\theta (\textbf{k})\right) & \frac{1}{2} \cos \theta (\textbf{k}) \partial_{\textbf{k}} \phi (\textbf{k}) \\
    \end{array}
    \right)}.
\end{eqnarray*}
\end{widetext}
This matrix is a vector-valued matrix. It means that the inter-band matrix elements, which relate the core with the valence and conduction bands, have out-of-plane components. In the other hand, the intra-band and inter-band matrix elements of the valence and conduction bands have in-plane components coming from the $\textbf{k}$-derivative.

The Berry curvature of the conduction state, the one took into account for the Saddle point approach and in the main text, is given by
\begin{equation*}
    \Omega(\textbf{K}) = \nabla_{\textbf{K}} \times \pmb{\xi}_{22}(\textbf{K}) = \frac{1}{2} \left( \nabla_{\textbf{K}} \cos\theta(\textbf{k}) \right) \times \left( \nabla_{\textbf{K}} \phi(\textbf{k}) \right)
\end{equation*}
\section{Semiclassical Model for the Absorption}
\label{sec:semiclassical}

We present here the derivation of Eq.~\ref{eq:polarization} for the polarization of the system. This expression is used to obtain a physical interpretation of first-principle calculations. We start deriving the light-induced dipole response of the system by using the following ansatz:
\begin{equation}
    \left| \Psi(t) \right> = \sum_{\textbf{K}} \left[b_{0}(\textbf{K},t) \left| g,\textbf{K} \right> + b_1(\textbf{K},t) \left| 1,\textbf{K} \right>  \right],
    \label{eq:ansatz}
\end{equation}
where $\left| g,\textbf{K} \right> = \hat{a}_{ch}^\dagger (\textbf{K}) \left| 0 \right>$ and $\left| 1,\textbf{K} \right> = \hat{a}_{c}^\dagger (\textbf{K}) \hat{a}_{ch} (\textbf{K}) \left| g,\textbf{K} \right>$, being $\hat{a}_{\alpha}^{\dagger}(\textbf{k})$ the operator that creates a particle with quasi-momentum $\textbf{k}$ in the $\alpha$ band. We are not considering excitations from valence to conduction because the difference in energy is too small for the electric field. \\
The time-dependent evolution of the amplitudes can be found by solving the time-dependent Schr\"odinger equation.
The Hamiltonian is written as $H(t) = H_0 + V_I (t)$, where $H_0$ is the unperturbed Hamiltonian, showed in the Appendix~\ref{sec:hamiltonian}, and $V_I (t )$ is the coupling between the electric field and the system in the length gauge. Assuming a sudden excitation at $t = t_0$ from the x-ray pulse (it couples the core with the conduction band), $\pmb{\varepsilon}_x (t) = \pmb{\varepsilon}_{x0} \delta(t - t_0 )$, then the evolution of the amplitudes is governed by
\begin{widetext}
\begin{align*}
    b_{0}(\textbf{K},t) &= 
    e^{ -i\int_{t_0}^{t} dt'\left[E_{0}( \textbf{K} +\textbf{A}(t') ) +\pmb{\varepsilon}_{IR}(t') \cdot \tilde{\pmb{\xi}}_{00}( \textbf{K} +\textbf{A}(t') )  \right] } b_{0}(\textbf{K},t_0), \\
    b_{1}(\textbf{K},t) &= 
    e^{ -i\int_{t_0}^{t} dt'\left[E_{1}( \textbf{K} +\textbf{A}(t') ) - i\Gamma_{ch}/2 +\pmb{\varepsilon}_{IR}(t') \cdot \tilde{\pmb{\xi}}_{11}( \textbf{K} +\textbf{A}(t') )  \right] } b_{1}(\textbf{K},t_0). \\
\end{align*}
\end{widetext}
Using the Haldane Hamiltonian and the Berry connection elements, we define $\tilde{\pmb{\xi}}_{00} = \pmb{\xi}_{00}$, $\tilde{\pmb{\xi}}_{11} = \pmb{\xi}_{22}+\pmb{\xi}_{00}$, $\tilde{\pmb{\xi}}_{01} = \pmb{\xi}_{02}$ and $E_0(\textbf{k}) = \epsilon_{ch}(\textbf{k})$, $E_1(\textbf{k}) = \epsilon_{c}(\textbf{k})$. $\epsilon_c$ and $\epsilon_{ch}$ are the energies for the conduction and core bands, respectively. $\Gamma_{ch}$ is the lifetime of the core-excited states and $\pmb{\varepsilon}_{IR}(t)$ is the external IR laser field. Using the Ansatz described in Eq,~(\ref{eq:ansatz}), the polarization can be written as 
\begin{equation}
    {\bf P}(t)  \propto  \sum_{{\bf K}}  \tilde{\pmb{\xi}}_{10}(\textbf{K}+\textbf{A}(t)) b_{1}^* (\textbf{K},t) b_{0}(\textbf{K},t) + c.c., 
\end{equation}
and this last expression can be rewritten as 

\begin{widetext}
\begin{equation}
    {\bf P}(t)  \propto  \sum_{{\bf K}} \left| \tilde{\pmb{\xi}}_{10} (\textbf{K}+\textbf{A}(t)) \right| 
    e^{-i S(\textbf{K},t,t_0)}
    e^{-\Gamma_{ch}(t-t_0)/2}   \;  b_{1}^* (\textbf{K},t_0) b_{0}(\textbf{K},t_0)
    + c.c., 
\end{equation}
\end{widetext}
where we define
 \begin{widetext}
 \begin{align*}
     \tilde{\pmb{\xi}}_{10}(\textbf{K}) &= \left| \tilde{\pmb{\xi}}_{10}(\textbf{K}) \right| e^{i\phi_{10}(\textbf{K})}, \\
      S (\textbf{K},t,t_0) &= \int_{t_0}^{t} dt' \left [ E_0(\textbf{K}') - E_1(\textbf{K}') + \pmb{\varepsilon}_{IR}(t')\cdot \tilde{\pmb{\xi}}_{00}(\textbf{K}') -  \pmb{\varepsilon}_{IR}(t')\cdot \tilde{\pmb{\xi}}_{11}(\textbf{K}') + \frac{d}{dt'} \phi_{10}(\textbf{K}') \right],\\
     &= -\int_{t_0}^{t} dt' \left[ \Delta E(\textbf{K}') + \pmb{\varepsilon}_{IR}(t')\cdot \Delta \tilde{\pmb{\xi}}(\textbf{K}')- \frac{d}{dt'} \phi_{10}(\textbf{K}') \right],
 \end{align*}
 \end{widetext}
and $\textbf{K}' = \textbf{K} + \textbf{A}(t')$. The last term is the gauge invariant semiclassical action $S (\textbf{K},t,t_0)$, in which the term $\Delta \tilde{\pmb{\xi}}(\textbf{K}) = \tilde{\pmb{\xi}}_{11}(\textbf{K}) - \tilde{\pmb{\xi}}_{00}(\textbf{K})$ involves the intra-band matrix elements of the Berry connection. Using the matrix elements of the Haldane model, the action takes the form
\begin{widetext}
\begin{equation}
    S(\textbf{K},t,t_0) = -\int_{t_0}^{t} dt' \left[  \epsilon_c(\textbf{K}') -\epsilon_{ch}(\textbf{K}') + \frac{1}{2} \pmb{\varepsilon}_{IR}(t') \cdot \partial_{\textbf{k}} \phi(\textbf{K}') \left[\cos\theta(\textbf{K}') +1  \right] \right].
\end{equation}
\end{widetext}

\section{Derivation of the Saddle Point Equations for the Polarization}
\label{sec:SPA}

We present here the derivation of Eqs. (\ref{eq:saddle_point_trajectory}) and (\ref{eq:saddle_point_energy})  of the main text, which are used to simplify the integrals over ${\bf k}$ obtained for the absorption within the semiclassical model. The SPA enables to solve easily an integral of the type
\begin{widetext}
\begin{eqnarray*}
    I_n(\sigma)=\int_C f(\mathbf{z}) e^{\sigma \varphi(\mathbf{z})} \mathrm{d} \mathbf{z} \approx \sum_s\left(\frac{2 \pi}{\sigma}\right)^{\frac{n}{2}} \frac{f\left(\mathbf{z}_s\right)}{\sqrt{\operatorname{det}\left(-\varphi_n^{\prime \prime}\left(\mathbf{z}_s\right)\right)}} e^{\sigma \varphi\left(\mathbf{z}_s\right)} =  \sum_s \sqrt{\frac{(2 \pi i)^n}{\operatorname{det}\left[\phi_n^{\prime \prime}\left(\mathbf{z}_s\right)\right]}} f\left(\mathbf{z}_s\right) e^{i \phi\left(\mathbf{z}_s\right)}
\end{eqnarray*}
\end{widetext}
where the points ${\bf z}_s$ are those that satisfy $\nabla_{\bf z} \varphi ({\bf z}) \vert_{\bf z_s} = 0$.

For our formalism, the integrals over ${\bf K}$ and time $t$ in the calculation of the dipole response in frequency domain can be reduced by SPA. For this, we calculate the critical points of the action, given by $\nabla_{\bf K} S(\textbf{K},t,t_0 ) = 0$ and $\partial_t S(\textbf{K},t,t_0 ) = 0$. For convenience, we re-write the action defined above as
\begin{widetext}
\begin{align*}
\hspace{-1cm}
    S(\textbf{K},t,t_0 ) = -\left [ \int_{t_0}^t \Big(\Delta E(\textbf{K}-{\bf A}(t)+\textbf{A}(t')) + \pmb{\varepsilon}_{IR} (t')\cdot \Delta \tilde{\pmb{\xi}}(\textbf{K}-{\bf A}(t)+\textbf{A}(t')) \Big) dt' \right. \\
    \left.  + \arg \Big(\tilde{\pmb{\xi}}_{10} (\textbf{K})\Big) - \arg\Big(\tilde{{\xi}}_{10}^\parallel\big(\textbf{K}-{\bf A}(t)+{\bf A}(t_0)\big)\Big) \right],
\end{align*}
\end{widetext}
by including the coupling between the core and the conduction band, on which we explicitly write the inter- and intra-band elements of the Berry connection and the dependence of each term with respect to the quasi-momentum $\textbf{K}$ and the IR vector potential at different times. The parallel component $\tilde{{\xi}}_{10}^\parallel$ is taken along the axis of polarization of the x-ray pulse. In this expression, $\arg \left( \tilde{\pmb{\xi}}_{10} \right)$ is the phase of the inter-band Berry connection matrix elements, i.e. $ \phi_{10} (\textbf{k})$. We define the semiclassical action as a vector ($\arg \left( \tilde{\pmb{\xi}}_{10} \right)$) meaning that to calculate the $\alpha$-th component of the polarization we need the $\alpha$-th component of the phase.


The critical points of the action must satisfy the equations
\begin{align*}
    & {\bf r}_d ({\bf K},t,t_0) - \pmb{\alpha}({\bf K}) + \pmb{\alpha}^\parallel (\textbf{K}-{\bf A}(t)+{\bf A}(t_0))= 0, \\
    &\epsilon_c({\bf k}) - \epsilon_{ch}({\bf k}) + \pmb{\varepsilon}_{IR} (t) \cdot  {\bf r}_d ({\bf K},t,t_0) + \\  &\qquad \, + \pmb{\varepsilon}_{IR} (t) \cdot \pmb{\alpha}^\parallel (\textbf{K}-{\bf A}(t)+{\bf A}(t_0))  = \omega,
\end{align*}
which are found by considering the conditions $\nabla_{\bf K} \left(\omega t +  S(\textbf{K},t,t_0 )\right) = 0 $ and $\partial_t\left(\omega t +  S(\textbf{K},t,t_0 ) \right)= 0$. In the previous equations we define ${\bf r}_d ({\bf K},t,t_0)$ as the relative distance between the created electron-hole pair
\begin{eqnarray*}
{\bf r}_d ({\bf K},t,t_0) = \int_{t_0}^t \Big( \nabla_{\bf K} \epsilon_c(\textbf{K}-{\bf A}(t)+\textbf{A}(t'))  \\
    + \bm{\varepsilon}_{IR} (t')\times \Omega (\textbf{K}-{\bf A}(t)+\textbf{A}(t'))  \Big) dt'.
\end{eqnarray*}
The terms inside the time integral can be interpreted as the instant velocity. The first term is the ballistic velocity that contains the derivative with respect to ${\bf k}$ of the energy dispersion, while the second term is the anomalous velocity that contains the Berry curvature of the conduction band $\Omega(\textbf{K}) = \nabla_{\textbf{K}} \times \pmb{\xi}_{22}(\textbf{K})$. Also, we define other variables, we call them Berry distances, that purely depend on the intra- and inter-band Berry connections
\begin{align*}
     \pmb{\alpha}({\bf K}) &= \bm{\xi}_{22}(\textbf{K}) - \nabla_{\bf K} [ \arg \Big(\bm{\xi}_{20} (\textbf{K})\Big) ],\\
    \pmb{\alpha}^\parallel({\bf K}) &= {\bm \xi}_{22} (\textbf{K}) - \nabla_{\bf K} [ \arg \Big({\pmb{\xi}}^\parallel_{20} (\textbf{K})\Big) ].
\end{align*}

These expressions have been obtained by using that $\nabla \left(\textbf{F}\cdot\textbf{G} \right) = \textbf{F}\times\left( \nabla \times \textbf{G} \right) + \textbf{G} \times \left( \nabla\times\textbf{F} \right) + \left(\textbf{F}\cdot\nabla  \right)\textbf{G} + \left(\textbf{G}\cdot\nabla  \right)\textbf{F}$. Note that that the Berry distances modify the condition for the recombination of the electron-hole pair. If those are zero, then we have the condition $ {\bf r}_d ({\bf K},t,t_0)=0$, and it can be interpreted as the creation of the electron-hole pair at $t=t_0$, when the attosecond pulse excites the system, and the propagation of both until their distance is zero. But the Berry structure of the material makes this process more complex, and the Berry distances modify this condition.

\end{appendices}
}
\begin{acknowledgments}

J.F.P.M, G.C., M.M., and A.P. acknowledge Comunidad de Madrid through TALENTO grant refs. 2017-T1/IND-5432 and 2021-5A/IND-20959, and the Spanish Ministry of Science, Innovation and Universities \& the State Research Agency through grants refs. PID2021-126560NB-I00 and CNS2022-135803 (MCIU/AEI/FEDER, UE), and the "Mar\'ia de Maeztu" Programme for Units of Excellence in R\&D   (CEX2023-001316-M), and FASLIGHT network (RED2022-134391-T), and computer resources and assistance provided by Centro de Computaci\'on Cient\'ifica de la Universidad Aut\'onoma de Madrid (FI-2021-1-0032), Instituto de Biocomputaci\'on y F\'isica de Sistemas Complejos de la Universidad de Zaragoza (FI-2020-3-0008), and Barcelona Supercomputing Center (FI-2020-1-0005, FI-2021-2-0023, FI-2021-3-0019). This publication is based upon work from COST Action NEXT, CA22148 supported by COST (European Cooperation in Science and Technology).  M. Malakhov's work also carried out within the state assignment of Ministry of Science and Higher Education of the Russian Federation (theme “Quantum” No. 122021000038-7).
E.P. acknowledges Royal Society funding under URF\textbackslash R1\textbackslash 211390, RF\textbackslash ERE\textbackslash 210255 and RF\textbackslash ERE\textbackslash 231081. 

ICFO and ex-ICFO co-authors acknowledge Europea Research Council AdG NOQIA; MCIN/AEI (PGC2018-0910.13039/501100011033,  CEX2019-000910-S/10.13039/501100011033, Plan National FIDEUA PID2019-106901GB-I00, Plan National STAMEENA PID2022-139099NB, I00, project funded by MCIN/AEI/10.13039/501100011033 and by the “European Union NextGenerationEU/PRTR" (PRTR-C17.I1), FPI); QUANTERA MAQS PCI2019-111828-2);  QUANTERA DYNAMITE PCI2022-132919, QuantERA II Programme co-funded by European Union’s Horizon 2020 program under Grant Agreement No 101017733); Ministry for Digital Transformation and of Civil Service of the Spanish Government through the QUANTUM ENIA project call - Quantum Spain project, and by the European Union through the Recovery, Transformation and Resilience Plan - NextGenerationEU within the framework of the Digital Spain 2026 Agenda; Fundació Cellex; Fundació Mir-Puig; Generalitat de Catalunya (European Social Fund FEDER and CERCA program, AGAUR Grant No. 2021 SGR 01452, QuantumCAT \ U16-011424, co-funded by ERDF Operational Program of Catalonia 2014-2020); 
Barcelona Supercomputing Center MareNostrum (FI-2023-3-0024); 
Funded by the European Union. Views and opinions expressed are, however, those of the author(s) only and do not necessarily reflect those of the European Union, European Commission, European Climate, Infrastructure and Environment Executive Agency (CINEA), or any other granting authority.  Neither the European Union nor any granting authority can be held responsible for them (HORIZON-CL4-2022-QUANTUM-02-SGA  PASQuanS2.1, 101113690, EU Horizon 2020 FET-OPEN OPTOlogic, Grant No 899794),  EU Horizon Europe Program (This project has received funding from the European Union’s Horizon Europe research and innovation program under grant agreement No 101080086 NeQSTGrant Agreement 101080086 — NeQST); ICFO Internal “QuantumGaudi” project; 
European Union’s Horizon 2020 program under the Marie Sklodowska-Curie grant agreement No 847648;  
“La Caixa” Junior Leaders fellowships, La Caixa” Foundation (ID 100010434): CF/BQ/PR23/11980043.

\end{acknowledgments}




\nocite{*}


\end{document}